\documentclass[prd,twocolumn,superscriptaddress,preprintnumbers,nofootinbib]{revtex4}
\usepackage{graphicx}
\usepackage{epsfig}
\usepackage{bm}
\usepackage{latexsym,amsmath,amssymb,wasysym,float}
\usepackage{mathrsfs}  
\usepackage{color}
\usepackage{tablefootnote}
\usepackage{enumitem}

\newcommand{\postscript}[2]{\setlength{\epsfxsize}{#2\hsize}
   \centerline{\epsfbox{#1}}}

\usepackage{float}
\usepackage{enumitem}
\usepackage{mathrsfs}  
\usepackage{hyperref}
\usepackage{multirow}

\usepackage[usenames,dvipsnames]{xcolor}
\definecolor{orange}{cmyk}{0,0.5,1,0}
\definecolor{rossoCP3}{cmyk}{0,.88,.77,.40}
\definecolor{graa}{rgb}{0.8,0.8,0.8}
\definecolor{blaa}{rgb}{0.2,0.2,0.6}

\def\simlt{\mathrel{\lower2.5pt\vbox{\lineskip=0pt\baselineskip=0pt
           \hbox{$<$}\hbox{$\sim$}}}}
\def\simgt{\mathrel{\lower2.5pt\vbox{\lineskip=0pt\baselineskip=0pt
           \hbox{$>$}\hbox{$\sim$}}}}

\newcommand{\be}{\begin{equation}}
	\newcommand{\ee}{\end{equation}}
\newcommand{\ba}{\begin{eqnarray}}
	\newcommand{\ea}{\end{eqnarray}}

 \allowdisplaybreaks

\newcommand{\nua}[1]{\ensuremath{\rlap
           {\kern-2.5pt\ensuremath
           {\overset{\scriptscriptstyle(-)}{\phantom{\nu}}}}
           {\ensuremath{{\nu}_{#1}}}}}

\begin{document}

\preprint{MPP-2024-199}
\preprint{LMU-ASC 17/24}

\title{\color{rossoCP3} SUSY at the FPF}

\author{\bf Luis A. Anchordoqui}

\affiliation{Department of Physics and Astronomy,  Lehman College, City University of
  New York, NY 10468, USA
}

\affiliation{Department of Physics,
 Graduate Center, City University
  of New York,  NY 10016, USA
}

\affiliation{Department of Astrophysics,
 American Museum of Natural History, NY
 10024, USA
}

\author{\bf Ignatios Antoniadis}

\affiliation{High Energy Physics Research Unit, Faculty of Science, Chulalongkorn University, Bangkok 1030, Thailand}

\affiliation{Laboratoire de Physique Th\'eorique et Hautes \'Energies - LPTHE, Sorbonne Universit\'e, CNRS, 4 Place Jussieu, 75005 Paris, France
}

\author{\bf Karim Benakli}

\affiliation{Laboratoire de Physique Th\'eorique et Hautes \'Energies - LPTHE, Sorbonne Universit\'e, CNRS, 4 Place Jussieu, 75005 Paris, France
}

\author{\bf Jules Cunat}

\affiliation{Laboratoire de Physique Th\'eorique et Hautes \'Energies - LPTHE, Sorbonne Universit\'e, CNRS, 4 Place Jussieu, 75005 Paris, France
}

\author{\bf Dieter L\"ust}

\affiliation{Max--Planck--Institut f\"ur Physik,  
 Werner--Heisenberg--Institut,
80805 M\"unchen, Germany
}

\affiliation{Arnold Sommerfeld Center for Theoretical Physics, \\
Ludwig-Maximilians-Universit\"at M\"unchen,
80333 M\"unchen, Germany
}

\begin{abstract}
\noindent Experimental searches for supersymmetry (SUSY) are entering a new
era. The failure to observe signals of
sparticle production at the
Large Hadron Collider (LHC) has eroded the central motivation for SUSY
breaking at the weak scale. However, String Theory requires SUSY at
the fundamental scale $M_s$ and hence SUSY could be broken at some high scale below $M_s$. Actually, if this were the case, the lack of
experimental evidence for
low-energy SUSY could have been
anticipated, because most stringy
models with high-scale SUSY breaking predict that sparticles would start popping up above about 10~TeV, well beyond the reach of current
LHC experiments. We show that using next generation LHC experiments currently envisioned for the
Forward Physics Facility (FPF) we could search for signals of
neutrino-modulino oscillations to probe models with string scale in
the grand unification region and SUSY breaking driven by sequestered gravity in gauge mediation. This is
possible because of the unprecedented flux of neutrinos to be produced as
secondary products in LHC collisions during the high-luminosity era and the capability of FPF
experiments to detect and identify their
flavors. 
\end{abstract}
\date{October 2024}
\maketitle

\section{Introduction}
  
It is well-known that supersymmetry (SUSY) provides a propitious
framework to alleviate the quadratic sensitivity of the Higgs sector
to unknown ultraviolet physics, which is unequivocally granted in the
Standard Model (SM) of particle physics. The most economic realization of
SUSY is the minimal supersymmetric standard model (MSSM) in which all
the SM fields have their supersymmetric partners. In the low energy
regime, SUSY is softly broken, and therefore the Higgs sector becomes
quadratically sensitive only to the sparticle soft masses. This
happenstance prompted weak scale sparticles to be the natural
candidates for steering the electroweak symmetry breaking. Therefore,
the search for sparticle signals has been considered as one of the
most compelling targets of experiments at the Large Hadron Collider (LHC).

Perhaps somewhat unexpectedly, LHC experiments have ruled out large
portions of the MSSM parameter space where 
SUSY would play a major role in solving the electroweak hierarchy
problem~\cite{ParticleDataGroup:2024cfk}. However, the SUSY saga is
still exhilarating because the 
regions excluded by LHC experiments correspond to minuscule portions
of the SUSY parameter space. Furthermore, SUSY is a natural and mostly
inevitable consequence of String Theory, the only consistent quantum
theory that includes gravity as well as the Yang-Mills force. Now, String Theory makes no requirements about the
masses of supersymmetric particles. Actually, in stringy models SUSY
is possibly broken at a very high scale.
If this were the case, it would be very challenging to observe signals
of superpartner production for existing LHC experiments.

In an accompanying Letter~\cite{Anchordoqui:2023qxv}, we proposed to probe models with high-scale SUSY
breaking by searching for neutrino disappearance  through oscillation
into modulinos at next generation LHC experiments currently envisioned for the
Forward Physics Facility (FPF). This is
possible because of the unprecedented flux of neutrinos to be produced as
secondary products in LHC collisions during the high-luminosity era and the capability of FPF
experiments to detect and identify their
flavors~\cite{Anchordoqui:2021ghd,Feng:2022inv}. Given typical
energies of LHC neutrinos $E \sim 100~{\rm GeV}$ propagating a
distance $L \sim 600~{\rm m}$, the modulino masses in the sensitivity
range of the FPF would be of the order of tens to hundreds of eV. As a natural
outgrowth of~\cite{Anchordoqui:2023qxv}, in this work we lay out the
foundations to pin down the SUSY breaking mechanisms that allow a
realization of our proposal.

The outline of the paper is as follows. We begin in Sec.~\ref{sec:2} by summarizing
the formalism of neutrino oscillations in the presence of a sterile
neutrino state (i.e. the modulino). Along the way, we also provide a
brief review of the sensitivity of the FPF experiments to the
neutrino-modulino oscillation phenomenon; for more details
see~\cite{Anchordoqui:2021ghd,Feng:2022inv,FASER:2019dxq,Bai:2020ukz}. In
Sec.~\ref{sec:3} we discuss constraints from cosmological observations. In
Sec.~\ref{sec:4} we reexamine a model  proposed
elsewhere~\cite{Antoniadis:2015chx} that has a string scale of the
order of the scale of grand unified theories (GUTs) and SUSY is broken
by a Scherk-Schwarz mechanism~\cite{Scherk:1978ta,Scherk:1979zr}. The SM gaugino, squark, and slepton masses arise predominantly from flavor blind gauge mediated interactions, but
the gravitino mass is superheavy due to an appropriate sequestering of the SUSY breaking
sector. We show that stringy models in which SUSY is broken via sequestered gravity in gauge mediation naturally
lead to a modulino that has a mass
within the discovery reach of the
FPF experiments. In Sec.~\ref{sec:5} we pay
attention to models with string scale  $M_s \sim 10^9~{\rm GeV}$ which
may be consistent with the dark
dimension, a five-dimensional
(5D) framework that has a compact space with characteristic
length-scale in the micron range~\cite{Montero:2022prj}. We first
summarize the main features of these stringy models and after that we
comment on some caveats and nuances of the modulino-mass estimate given in~\cite{Anchordoqui:2023qxv}. The
paper wraps up in Sec.~\ref{sec:6} with some conclusions.

\section{Neutrino-Modulino Oscillations}
\label{sec:2}

It has long been recognized that Planck
 suppressed interactions could
be connected to neutrino physics, because the coupling $M_p^{-1} LLHH$
generates a mass $\langle H \rangle^2/M_p \sim 10^{-1}~{\rm meV}$ near
the neutrino mass scale,
where  $M_p \simeq 2.48 \times 10^{18}~{\rm GeV}$ is the reduced Planck
mass, $L$ is the leptonic doublet and $H$ the Higgs doublet of the SM~\cite{Barbieri:1979hc,Akhmedov:1992hh}. It has also been recognized that in String Theory the superpartners of moduli
fields yield feasible candidates for the right-handed
neutrinos~\cite{Lukas:2000wn,Lukas:2000rg}. Before SUSY breaking, some of these
moduli fields as well as their fermionic partners are exactly
massless. Masses for the moduli and modulinos are then generated 
when SUSY is broken and can be small. This may account for the
lightness of any sterile neutrino
states.

Herein, we generalize the
idea introduced in~\cite{Benakli:1997iu} and assume that among the modulinos there is at
least one, $\tilde{z}$ (associated to modulus $z$), with the following properties:
\begin{itemize}[noitemsep,topsep=0pt]
\item $\tilde{z}$ has only Planck mass suppressed interactions with SM fields, as expected for geometrical moduli governing the different
  couplings between light fields. We work within a simple construct,
  in which  the relevant light scale  for SM singlets is the gravitino
  mass $m_{3/2}$ and a dimensionless coupling constant with an active SM
  neutrino is given by 
\begin{equation}
\lambda_\alpha =\alpha_\alpha \ \frac{m_{3/2}}{M_p} \, , 
\label{lambda_i}
\end{equation}
where $\alpha_\alpha$ is an additional  suppression factor that
 parametrizes our ignorance of the 
 interaction between the visible and
 hidden sectors of particular models, and the subindex $\alpha =
 e,\mu,\tau$ indicates the neutrino flavor.
 \item The mixing of $\tilde{z}$ with the active
SM neutrinos involves the electroweak symmetry breaking,
and the simplest appropriate effective operator is $\lambda_\alpha \overline L_\alpha
\tilde{z} H$. This operator generates a neutrino mass term
$m_{\nu_\alpha \tilde z} \bar \nu_\alpha
\tilde{z}$ with  
\begin{equation}
m_{\nu_\alpha \tilde{z}} = \alpha_\alpha  \ \eta \ \frac{m_{3/2} \langle H \rangle}{M_p} \, ,
\label{mnu}
\end{equation}
where $\eta$ parametrizes the renormalization effect.
\item The mass of the modulino, $m_{\tilde z}$, is induced via
  SUSY breaking. We assume that $m_{\tilde z}$ is absent at the level
  $m_{3/2}$ and appears as
  \begin{equation}
    m_{\tilde z} = \beta \ \frac{m_{3/2}^{k+1}}{M_p^k} \,,
\label{m4}
  \end{equation}
where $\beta$ is a model dependent suppression factor and $k \in
\mathbb{N}^+$.
\end{itemize}
In the Secs.~\ref{sec:4} and \ref{sec:5} we investigate models that provide specific
realizations of these assumptions for $k=2$ and $k=1$, respectively.

Neutrino-modulino oscillations are equivalent to the disappearance
phenomenon governed by neutrino oscillations
in a (3 + 1) framework, in which the flavor states
are given by the superposition of four massive neutrino states,
\begin{equation}
  |\nu_\alpha \rangle = \sum_{i =1}^4 U^*_{\alpha i} \ | \nu_i\rangle \,,
\end{equation}
where $\alpha = e,\mu, \tau, \tilde z$~\cite{Dasgupta:2021ies}. The presence of the modulino reshapes the active neutrino mixing angles via the unitarity
relations of the mixing matrix $\mathbb U$, i.e.,
\begin{equation}
\sum_{\alpha} U^*_{\alpha i} \ U_{\alpha j} = \delta_{ij} \quad \quad {\rm
  and} \quad \quad
  \sum_{i=1}^4 U^*_{\alpha i} \ U_{\beta i} = \delta_{\alpha \beta} \, ,
\label{uni2}
\end{equation}
where Greek indices run over the neutrino flavors and Roman indices
over the mass states, with $m_4 \equiv m_{\tilde z}$.
Bearing this in mind, the unitary $4 \times 4$ mixing matrix takes the form 
$\mathbb U = \mathbb U_{34} \ \mathbb U_{24} \ \mathbb U_{14} \
\mathbb U_{\rm
  PMNS}$
where $\mathbb U_{\rm PMNS} = \mathbb U_{23} \,
\mathbb U_{13} \, \mathbb U_{12}$ is the the Pontecorvo-Maki-Nagakawa-Sakata (PMNS)
matrix~\cite{Pontecorvo:1967fh,Pontecorvo:1957qd,Maki:1962mu}.

Data analyses from short- and long-baseline neutrino oscillation
experiments, together with observations of neutrinos produced by
cosmic rays collisions in the atmosphere and by nuclear fusion
reactions in the Sun, provide the most sensitive insights to determine
the extremely small mass-squared differences. Neutrino oscillation data can
be well-fitted in terms of two nonzero differences $\Delta
m^2_{ij}=m^2_i-m^2_j$ between the squares of the masses of
the three ($i=1,2,3$) neutrino mass eigenvalues $m_i$, yielding $\Delta m_{\rm
  SOL}^2 = \Delta m^2_{21} = (7.53 \pm 0.18) \times 10^{-5}~{\rm
  eV}^2$  and $\Delta m^2_{\rm ATM} = |\Delta m_{31}^2| \simeq \Delta m^2_{32} = 2.453 \pm 0.033) \times 10^{-3}~{\rm
  eV}^2$ or $\Delta m^2_{32} = -2.536 \pm 0.034) \times 10^{-3}~{\rm
  eV}^2$~\cite{ParticleDataGroup:2024cfk}.

The disappearance phenomenon is driven by the Hamiltonian
\begin{equation}
  {\cal H} = \frac{1}{2E} \mathbb{U} \mathbb{M}^2 \mathbb{U}^\dagger
\end{equation}  
where $\mathbb{M}^2 \equiv {\rm diag} (0, \Delta m_{21}^2, \Delta
m_{31}^2, \Delta m_{41}^2)$, and where 
$\Delta m_{41}^2$ is the modulino-neutrino mass squared difference. The $\nu_{\alpha} \to \nu_{\beta}$ transition probability is found to be
\begin{equation}
  P_{\alpha \beta} = \left|\sum_{i=1}^4 = U^*_{\alpha
      i} \ U_{\beta
      i} \ \exp\left(-i \frac{\Delta m_{i1}^2 \ L}{2E} \right) \right|^2 \, ,
\end{equation}
where $L$ is the experiment baseline.  In the FPF-short-baseline limit
({\it viz.}, $\Delta m^2_{21}L/E \ll 1$ and $\Delta m^2_{31}L/E \ll 1$) for which SM oscillations have not
developed yet, the effective oscillation probabilities
can be written as 
\begin{equation}
 \!\!\!  P^{\rm FPF}_{\alpha \beta}  = \left|\sum_{i=1}^3 U^*_{\alpha i}
    U_{\beta i} + U^*_{\alpha 4}  U_{\beta 4}  \exp \left(-i
      \frac{\Delta m_{41}^2 L}{2E} \right)\right|^2 \! \! .
\label{tp1}  
\end{equation}
Using (\ref{uni2}), the transition probability (\ref{tp1}) can be recast as
\begin{eqnarray}
  P^{\rm FPF}_{\alpha \beta}  & = &  \left| \delta_{\alpha \beta} -
  U^*_{\alpha 4} \ U_{\beta 4} \left[ 1 - \exp\left(-i \frac{\Delta
                                                           m^2_{41} \ L}{2E} \right)\right]
                                                           \right|^2
                                                           \nonumber \\
& = &  \delta_{\alpha \beta} - \sin^2 2 \theta_{\alpha \beta} \ \sin^2  \left( \frac{\Delta m^2_{41}
     \  L}{4E} \right)  \, ,
\end{eqnarray}
where $\sin^2 2 \theta_{\alpha \beta} = 4 \ |U_{\alpha 4}|^2  \left(\delta_{\alpha
       \beta} - |U_{\beta 4}|^2\right)$. The disappearance phenomenon
   is then driven by the oscillation probability 
\begin{equation}
  P_{\alpha \alpha}^{\rm FPF} = 1 - 4 |U_{\alpha 4}|^2
  (1-|U_{\alpha 4}|^2) \sin^2 \left(\frac{\Delta m_{41}
      L}{4E}\right)\, .
\end{equation}

\begin{figure}[tb]
  \postscript{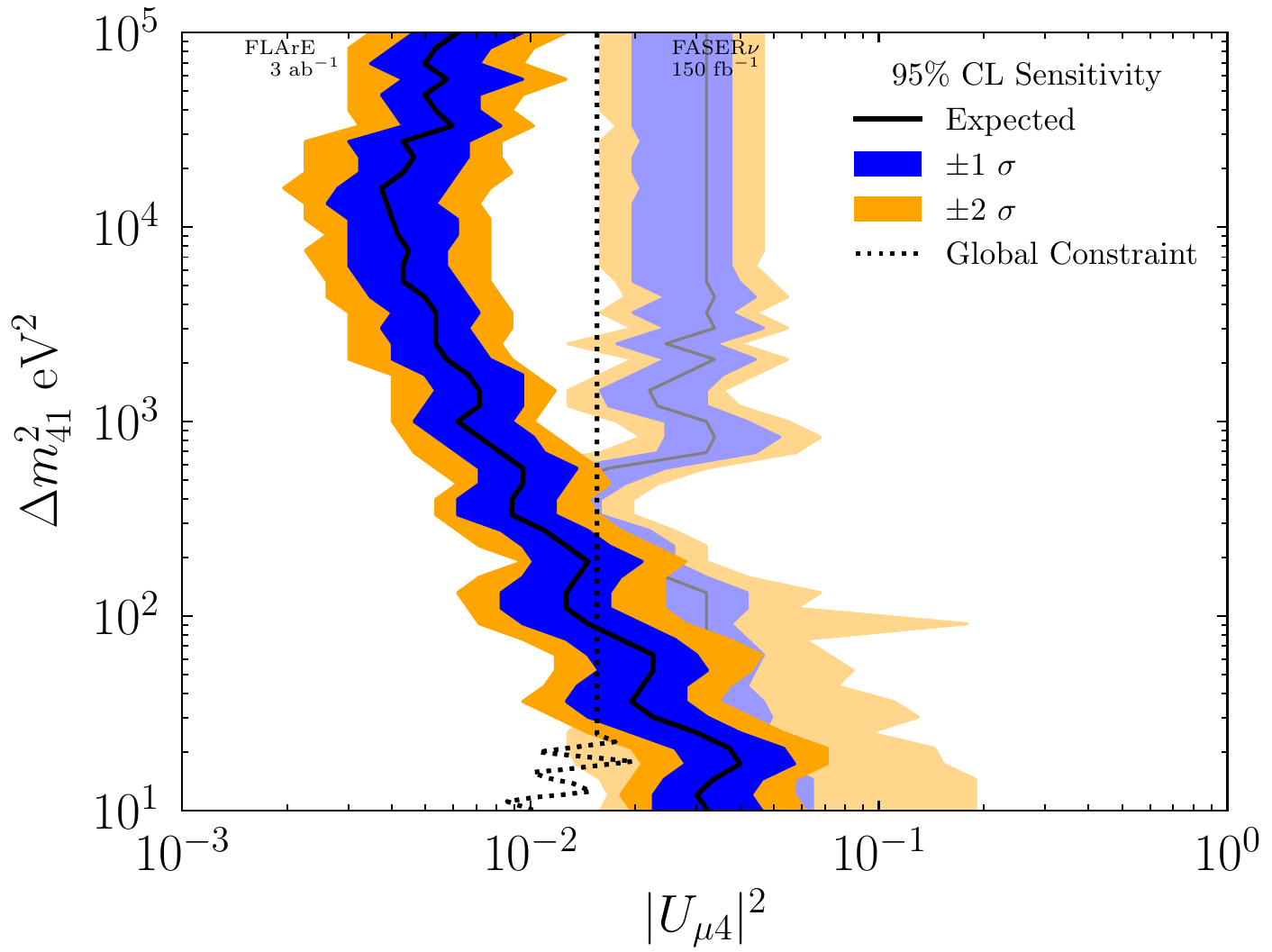}{0.9}
  \caption{Projected determinations of the mixing parameter $|U_{\mu 4}|^2$ for: {\it (i)} FASER$\nu$ at LHC Run 3 with
    $150~{\rm fb^{-1}}$ (light shaded regions) and {\it (ii)}~FLArE at the
    Hight Luminosity LHC with $3~{\rm ab}^{-1}$ (dark shaded
    regions); 68\% and 95\% CL
    exclusion contours are shown for both experiments. The contours have been derived using
    the Feldman-Cousins procedure~\cite{Feldman:1997qc}. Constraints
    on the allowed mixing strength $|U_{\mu 4}|^2$ from a global
  (3+1) analysis of neutrino oscillations~\cite{Dentler:2018sju} are
  shown for comparison. From Ref.~\cite{Anchordoqui:2021ghd}. \label{fig:1}}
\end{figure}  

In Fig.~\ref{fig:1} we show the $(|U_{\mu 4}|^2, \Delta m_{41}^2)$
parameter space that can be probed by FASER$\nu$ at LHC Run 3 and by FPF's
FLArE during the LHC high-luminosity era. It is evident that FLArE
will have a leading discovery reach for modulino in the $m_4 \agt 10~{\rm
  eV}$ mass region.  

\section{Constraints from
  Cosmology}
\label{sec:3}

The additional radiation energy provided by the modulino could modify
the cosmic expansion rate and affect: {\it (i)}~the big-bang nucleosynthesis
(BBN) of light elements, {\it (ii)}~the  anisotropies imprinted in
the cosmic microwave background (CMB),
and the formation of large-scale structures (LSS). In this section we
examine the compatibility of a light modulino with recent cosmological observations.

\begin{figure}[tb]
  \postscript{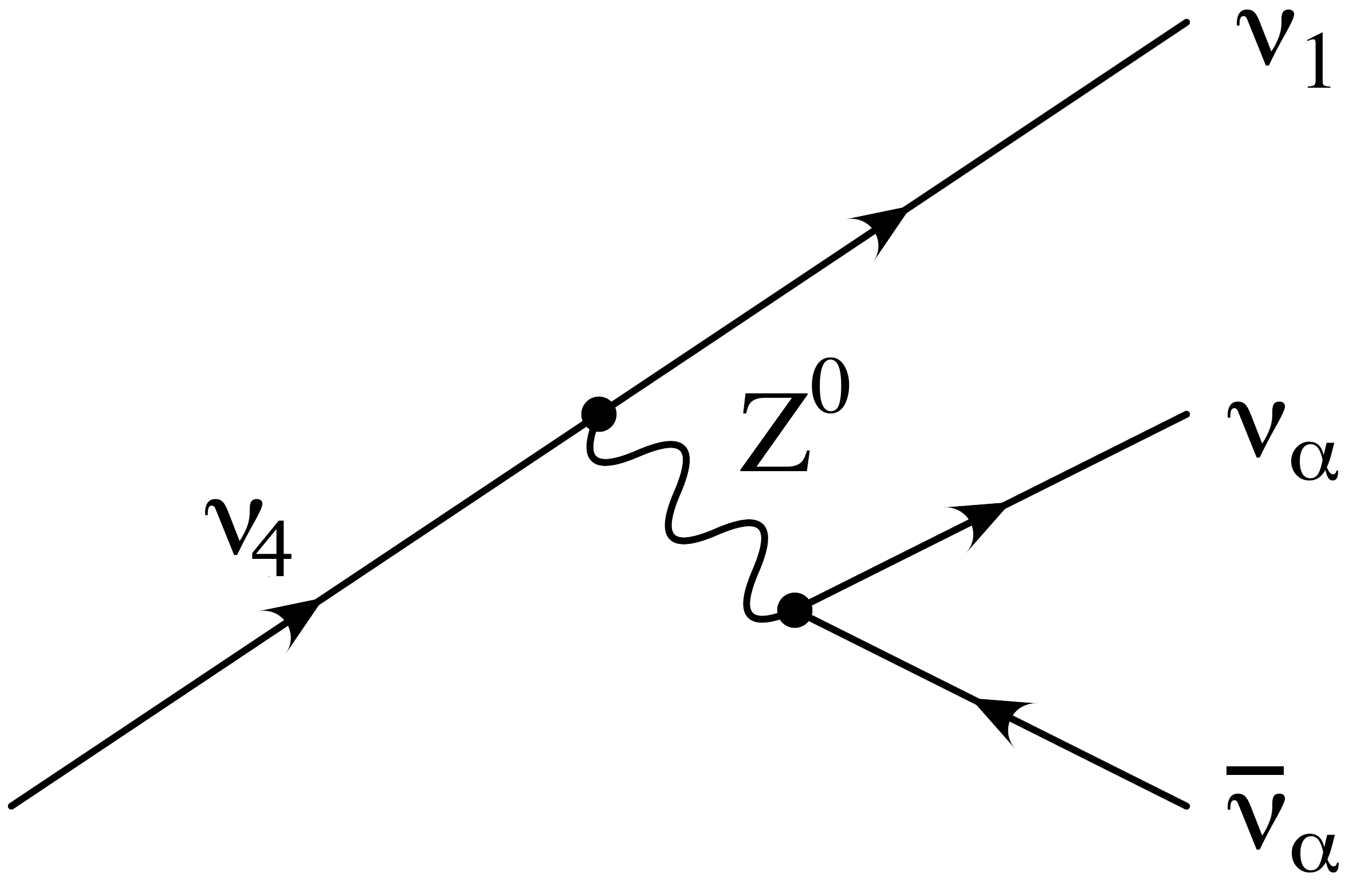}{0.7}
  \caption{Dominant decay mode for a modulino with mass less
    than twice the electron mass. There are three light active
    neutrinos in the final state.
\label{fig:a}}
\end{figure}

\begin{figure*}[tb]
  \postscript{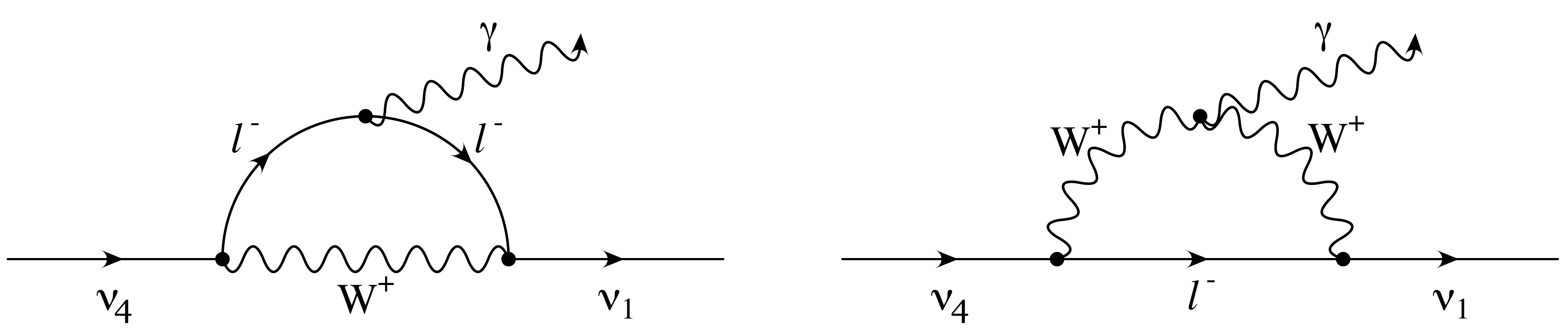}{0.8}
  \caption{Radiative decay modes for massive sterile neutrinos. Majorana fermions also have contributions from conjugate processes~\cite{Pal:1981rm}.
\label{fig:b}}
\end{figure*}

The modulino has finite
lifetime. The primary decay channel of a modulino with a mass
below twice the electron mass is into three light active neutrinos, as
shown in Fig.~\ref{fig:a}~\cite{Dicus:1977qy}.  A weak interaction decay width is usually
assumed to scale as the mass to the fifth power because the weak
coupling constant has dimensions of inverse mass squared. For the
diagram of Fig.~\ref{fig:a}, the partial decay width is dressed by the
active-sterile mixing angle $\theta_{\alpha \tilde z}$ and so the decay rate is found to be 
\begin{eqnarray}
\Gamma_{3\nu} &\approx&  G_{\rm F}^2 \ \sin^2 (2\theta_{\alpha \tilde z})
\ \left(\frac{m_4^5}{768\pi^3}\right) \nonumber \\ &\approx& 8.7\times 10^{-28}
\left(\frac{\sin^2 2\theta_{\alpha \tilde z}}{10^{-2}}\right)
\left(\frac{m_4}{100~{\rm\, eV}}\right)^5~{\rm s^{-1}} ,
\end{eqnarray}
where $G_F \approx 10^{-5}/m_p^2$ is the Fermi
constant~\cite{Abazajian:2000hw,Abazajian:2001vt}. Note that for the mass range of interest, the modulino is
quasi-stable because its lifetime is excidingly larger than the age of
the universe, which is ${\cal O} (10^{17}~{\rm s})$.
The dominant radiative decay modes are shown in Fig.~\ref{fig:b}.  The partial
width of the radiative decay  $\nu_4 \to\gamma \nu$ is smaller than
that of $\nu_4 \to 3\nu$ by a
factor of $27 \alpha/(8\pi)$, with $\alpha$ the fine structure
constant~\cite{Barger:1995ty}. All in all, we conclude that for the
mass range of interest, these
processes have very little impact on LSS formation.

In the early universe, flavor oscillations would bring the sterile
state into thermal equilibrium prior to neutrino decoupling (at
temperature $\sim 1~{\rm MeV}$),
increasing the relativistic energy density.  The degree of
thermalization of the sterile state can be expressed in terms of the effective number
of relativistic neutrino-like
species $N_{\rm eff}$~\cite{Steigman:1977kc}, which is bounded by
experiment~\cite{Planck:2018vyg,Steigman:2012ve}. For a (3+1) mixing, the sterile
neutrino contributes with $\Delta N_{\rm eff}$. The degree of
thermalization $\Delta N_{\rm eff}$ is connected to the mixing
parameters by
\begin{equation}
\frac{\Delta m_{41}^2}{\rm eV} \ \sin^4 2 \theta_{\alpha \tilde z}= 10^{-5} \ \ln^2 ( 1- \Delta
N_{\rm eff}) \, ,
\end{equation}
with $\sin^2 2\theta_{\alpha \tilde z} \simeq 4 |U_{\alpha 4}|^2
|U_{\tilde z 4}|^2$~\cite{Hagstotz:2020ukm}. As shown in Fig.~1
of~\cite{Hagstotz:2020ukm} in the $(\Delta m_{41}^2, U_{\mu
  4})$ range of to be probe by the FPF
experiments, the modulino contribution
to $\Delta N_{\rm eff}$ 
is in tension with the upper limit reported by the Planck Collaboration~\cite{Planck:2018vyg}. 

Over the last few years, low- and high-redshift observations set off
tensions in the measurement of some cosmological
parameters~\cite{Abdalla:2022yfr}. Obviously, the CMB bound on $\Delta
N_{\rm
  eff}$ cannot be taken as written in stone until the the cosmological
tensions are resolved~\cite{Lesgourgues}.  If {\it Planck}'s CMB constraint were
to be removed, it is interesting to note that the large mixing matrix elements to be probed
at the FPF experiments increase $\Delta N_{\rm eff}$ towards one, a
value consistent with BBN
predictions~\cite{Steigman:2012ve}. Alternatively, following~\cite{Abazajian:2023reo} one can
extend the particle spectrum in the far infrared so that 
the modulino could decay into a lighter modulino and a scalar. If this
were the case, then the modulino contribution to $\Delta N_{\rm eff}$
would be ameliorated to accommodate {\it Planck}'s CMB constraint. In what follows we proceed on the assumption of any of these two premises.

\section{SUSY breaking with Sequestered gravity in Gauge Mediation}
\label{sec:4}

A fundamental controversy of supersymmetric models is the understanding
of SUSY breaking. Supergravity theories come up with a straightforward
pathway in which SUSY is broken by an $F$-term of a hidden superfield,
and gravity is the mediator that transmits SUSY breaking to the
ordinary quarks and leptons~\cite{Barbieri:1982eh,Chamseddine:1982jx}. If SUSY
is broken by a non-zero $F$-term, then the gravitino acquires a Majorana mass
$m_{3/2} \neq 0$ and it is generally expected that modulino masses
would be of the same 
order of the gravitino mass. To suppress the modulino mass we must
isolate the {\it gravitino sector} from the {\it modulino sector}. On
account of this, herein we consider the mechanism of SUSY breaking put forward elsewhere~\cite{Antoniadis:2015chx}, which is
embeddable in String Theory and simultaneously shares the main
advantages of (sequestered) gravity and gauge mediation.

To develop
our program in the simplest way, we work within the construct of
a minimal model in which the MSSM is localized in effectively three (spatial)
dimensions, on a collection of D-brane stacks, transverse to an extra dimension $y$ on a
semi-circle (orbifold) of radius $R$,
along which there is a bulk hidden gauge group $G_H$. We assume that
the hidden group has a non-chiral spectrum. Matter fields are described by excitations of open strings stretched between the SM brane and the $G_H$ D-brane extended in the
bulk, and hence they are localized in their three-dimensional (spatial)
intersection. These matter  
fields transform in the corresponding bi-fundamental
representations, and since we assumed that they are non-chiral, they can get a
mass by appropriate brane displacements (or analogously Wilson
lines), that we take as free model parameter. Moreover, these
localized fields are responsible for transferring SUSY breaking to the observable sector.

SUSY breaking is induced by a Scherk-Schwarz~\cite{Scherk:1978ta,Scherk:1979zr} deformation
along the $y$ dimension which generates a Majorana mass for the gravitino proportional to
the compactification scale $1/R$, but leaves the SM brane
supersymmetric~\cite{Antoniadis:1998ki}. 
The breaking is mediated to the observable SM sector by
both gravitational~\cite{Antoniadis:1998ki,Antoniadis:1997ic} and gauge interactions~\cite{Benakli:1998pw} (through the
bi-fundamental messenger fields), whose relative strength is
controlled by the compactification scale and messenger mass. The
Higgsino mass $\mu$ and soft Higgs-bilinear $B_\mu$-term could be
generated at the same order of magnitude as the other soft terms by
effective supergravity couplings as in the Giudice-Masiero mechanism~\cite{Giudice:1988yz}. Fixing
for definiteness the MSSM soft terms at the TeV scale and requiring
the gravitational contribution to be suppressed with respect to the
gauge mediated one by at least one order of magnitude, it follows that
the compactification scale 
\begin{equation}
10^{11.0} \alt \frac{1}{R~{\rm GeV}} \alt 10^{11.5} \,,
\label{yRscale}
\end{equation}
corresponding
to $M_s
\sim M_{\rm GUT} \sim 10^{16}~{\rm GeV}$,which is inferred by extrapolating the low energy SM gauge couplings with SUSY~\cite{Antoniadis:2015chx}.
 
Thus, critically for our purposes, the {\it modulino sector} has been isolated
from the {\it gravitino sector}. The gravitino is in the bulk and
acquires its mass at tree level, whereas the modulino is localized on the
boundary (e.g., at $y =0$) and hence $m_{\tilde z} = 0$ at tree
level.  The modulino mass is generated by radiative corrections. If we
consider moduli with gravitational couplings then the mass is
generated by gravitational radiative corrections. We do not further
discuss here the details of the messenger or observable sectors and
point the reader to~\cite{Antoniadis:2015chx} for details. Instead we
pay attention to the calculation of the modulino mass.

\begin{figure}[tb]
  \postscript{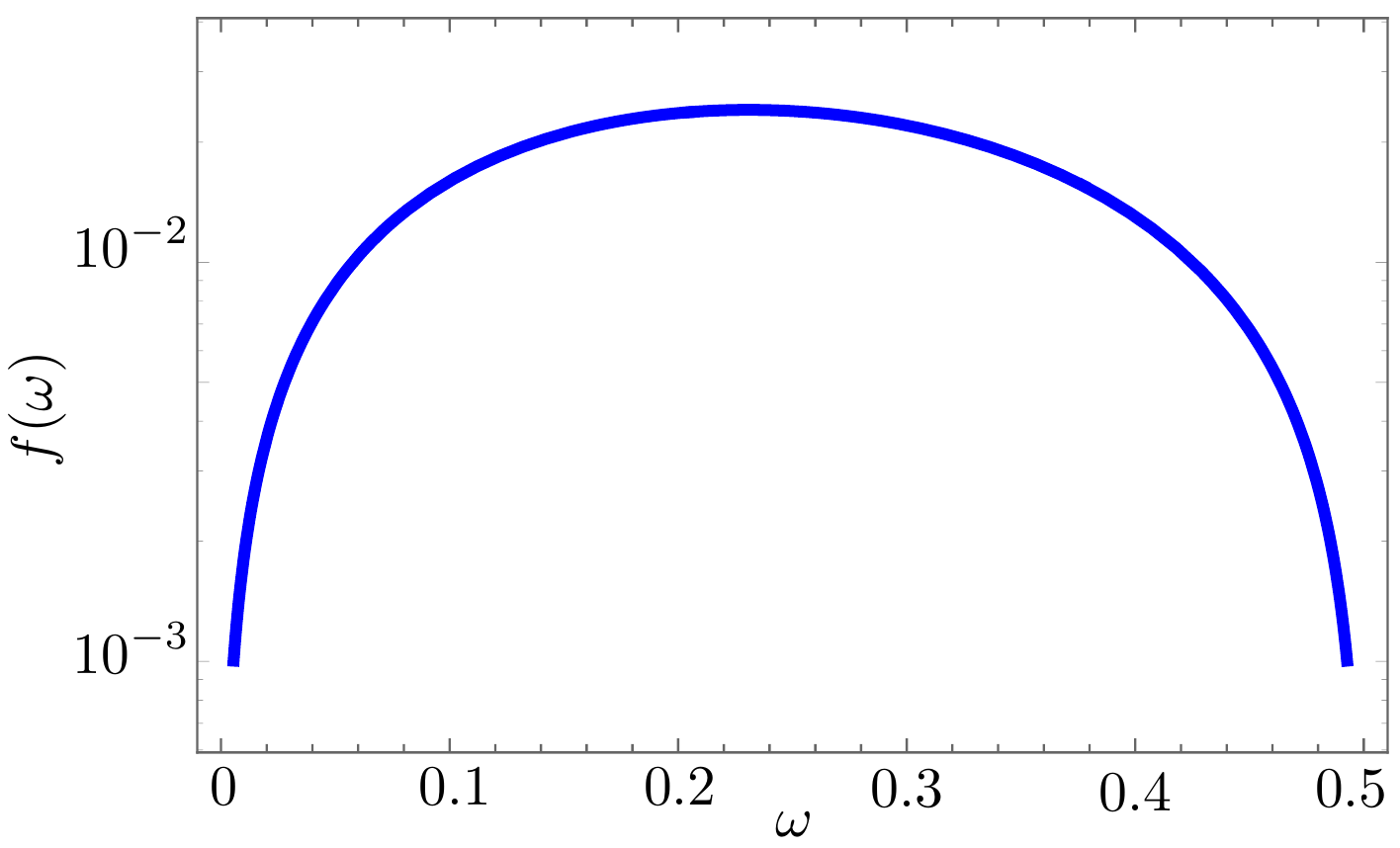}{0.9}
  \caption{The function $f (\omega)$. \label{fig:2}}
\end{figure}

Since we are envisioning SUSY to be broken via a Scherk-Schwarz mechanism,
the gravitino Kaluza-Klein (KK) modes get a mass
\begin{equation}
M_{n}(\omega)=m_{3/2} + \frac{n}{R},
\label{m32}
\end{equation}
where $m_{3/2} = \omega/R$ is the mass of the gravitino zero mode, $n
\in \mathbb N$, and $\omega$ is a
real parameter $0<\omega<\frac{1}{2}$. SUSY
breaking is transmitted from the bulk to the brane by one-loop
gravitational interactions giving a Majorana mass to the modulino,
which is given by 
\begin{eqnarray}
m_{\tilde z} & = & \frac{1}{M_p^2}\sum_n\int
                   \frac{d^4k}{(2\pi)^4}\left[\frac{M_n(\omega)}{k^2+M_n^2(\omega)}-\frac{M_n(0)}{k^2+M_n^2(0)}\right]
                   \nonumber \\
 & = & \frac{1}{R}\frac{1}{(M_pR)^2}\frac{1}{(4\pi)^2}
                   f(\omega) 
  \label{modumass}
\end{eqnarray}
where
\be
f(\omega)=\frac{3i}{8\pi^3}\left[{\rm
    Li}_4(e^{-2i\pi\omega})-{\rm Li}_4(e^{2i\pi\omega})
\right] \,,
\ee
with ${\rm Li}_s(z) = \sum_k z^k/k^s$  the polilogarithm and $\zeta
(x)$ the Riemann zeta function~\cite{Antoniadis:2015chx,Antoniadis:1997ic,Gherghetta:2001sa}. In Fig.~\ref{fig:2} we show
the function $f(\omega)$. Note that because of a preserved $R$-symmetry, $f(\omega)$ vanishes for $\omega = 1/2$~\cite{Antoniadis:2015chx,Antoniadis:1997ic,Gherghetta:2001sa}.

We now turn to relate the modulino mass
given in (\ref{modumass}) to the discovery reach of the FPF
experiments. To this end we concentrate on the region of the
parameter space with $0.1<\omega <0.4$, for which $0.01 < f(\omega) <
0.02$. Substituting this range of $f(\omega)$ together with the
benchmark compactification scale (\ref{yRscale}) into  (\ref{modumass}) we obtain
\begin{equation}
10 \alt \frac{m_{\tilde
  z}}{{\rm eV}}
  \alt 500 \, .
\end{equation}
This corresponds to
\begin{equation}
  10^2 \alt \frac{\Delta m_{41}^2}{{\rm eV}^2} \alt 10^5 \,,
\end{equation}
a range that spans the FPF discovery reach shown in Fig.~\ref{fig:1}.

A definite observation of $\tilde z$ signals at the FPF experiments
depends on the mixing angle $\theta_{\alpha \tilde{z}}$, with
\begin{equation}
  \tan 2 \theta_{\alpha \tilde z} = \frac{m_{\nu_\alpha \tilde{z}}}{m_4 - m_{\nu_i}} \,,
\end{equation}
and where $m_{\nu_i} \ll m_{4}$ are the neutrino mass states, $\alpha
= e,\mu,\tau$, and $i=1,2,3$~\cite{Benakli:1997iu}. Note that exisitng
data from neutrino oscillation experiments in general and future FPF
data in particular could provide a constraint on the model dependent product of
$\alpha_\alpha$ times $\eta$. 

In closing, we note that the
Search for Hidden Particles (SHiP) experiment would attain a comparable
sensitivity to the mixing between sterile and tau
neutrinos~\cite{Choi:2024ips}. This
would allow for a simultaneous, completely independent, verification of
neutrino-modulino oscillation signals.

\section{Lessons from the Swampland}
\label{sec:5}

The Swampland program seeks to characterize the space of effective
field theories (EFTs) coupled to gravity with a consistent UV
completion~\cite{Vafa:2005ui}. Such a set of EFTs are placed in the landscape (i.e. the area
compatible with quantum gravity) while other EFTs are usually placed
in an area called the swampland, which is incompatible with quantum
gravity. It goes without saying that the swampland is wider than the
landscape, and actually it
surrounds the landscape. There are several conjectures for fencing off
the swampland~\cite{Palti:2019pca,vanBeest:2021lhn,Agmon:2022thq,Anchordoqui:2024rov}. For instance, the distance conjecture (DC)
advocates that when venturing to large distances in 4D Planck units
within scalar field space of any consistent theory of quantum gravity,
a tower of particles will become light at a rate that is exponential
in the field space distance~\cite{Ooguri:2006in}. Associated to the DC
is the anti-de Sitter (AdS) distance conjecture, which relates the
dark energy density to the mass scale $m$ characterizing the infinite
tower of states, $m \sim |\Lambda|^\alpha$, as the negative AdS vacuum energy
 $\Lambda \to 0$, with $\alpha$ a positive constant of ${\cal O}
 (1)$~\cite{Lust:2019zwm}. Actually, if this scaling behavior stays
 around valid in de Sitter (or
 quasi de Sitter) space, an unbounded number of massless modes would also emerge in the limit $\Lambda \to 0$.
 
Strikingly, one can draw a connection between the AdS-DC in de Sitter
(dS) 
space and experimental observations on deviations from Newton's
gravitational inverse-square law~\cite{Lee:2020zjt} as well as neutron
star heating~\cite{Hannestad:2003yd} to elucidate the cosmological hierarchy
 problem: $\Lambda/M_p^4 \sim 10^{-120}$.
Indeed, the
smallness of the dark energy $\Lambda$ in Planck units could be understood as
an asymptotic limit in field space, corresponding to a
decompactification of one extra dimension of a size in the
micron range~\cite{Montero:2022prj}. More concretely, this dark
dimension opens up at the characteristic mass scale of the KK tower,
\begin{equation}
m_{{\rm KK}_1} \sim \lambda^{-1} \ \Lambda^{1/4} \,,
\label{mkk1}
\end{equation}
where $10^{-2} \alt \lambda \alt
10^{-4}$.\footnote{The dark dimension can also be viewed as a line interval with end- of-the-world 9-branes attached at each end~\cite{Schwarz:2024tet}.}  The
5D Planck scale (or species
scale where gravity becomes strong) is
given by
\begin{equation}
 M_* \sim \sqrt{\frac{8\pi}{N}} \ M_p \sim \sqrt{8\pi} \
 \lambda^{-1/3} \ \Lambda^{1/12} \ M_p^{2/3} \,, 
\end{equation}
where $N$ is the number of light species~\cite{Dvali:2007hz,Dvali:2007wp}. As a matter of course, SM fields are confined to 3-branes which are
localized in a higher-dimensional gravitation-only bulk~\cite{Antoniadis:1998ig}. Thus, bulk
fermions are SM singlets, which can be connected to KK towers of 4D
sterile neutrinos. In this direction, the dark dimension lays out a
cost-effective framework to realize an old idea for explaining the
smallness of neutrino masses by introducing the
right-handed neutrinos as 5D bulk states with Yukawa couplings to the left-handed lepton and Higgs doublets that are localized states on the SM brane stack~\cite{Dienes:1998sb,Arkani-Hamed:1998wuz,Dvali:1999cn}. The neutrino masses are then suppressed due to the wave function of the bulk states.

Without further ado we examine the relation between the SUSY breaking scale
and the measured value of the dark energy density $\Lambda$. The first
step towards associating $m_{3/2}$ to the mass scale of an infinite
tower of KK modes was taken in~\cite{Antoniadis:1988jn}, and this
idea has been recently formulated as the gravitino conjecture~\cite{Cribiori:2021gbf,Castellano:2021yye}. The
dark dimension and the gravitino conjecture give rise to two
possible schemes, depending on the relation between the
corresponding towers of external states~\cite{Anchordoqui:2023oqm}. A first possibility is that $\Lambda$ and $m_{3/2}$ are connected to
the same KK tower. A second possibility is that the towers are
different.

Herein we focus attention on the second possibility and assume that together with the infinite tower associated to the dark dimension there is a second tower of KK states related to $p$
 additional compact dimensions, with mass scale $m_{{\rm KK}_2}$. \footnote{As shown in~\cite{Anchordoqui:2023oqm} the scenario in which $m_{3/2}$ is related to the KK modes of the
micron-size dimension has associated very light gravitinos.} In this
case, the
quantum gravity cut-off is given by~\cite{Anchordoqui:2023oqm}
\begin{equation}
M_* = (8 \pi)^{2/(3+p)} \ m_{{\rm KK}_1}^{1/(3+p)} \ m_{{\rm
    KK}_2}^{p/(3+p)} \
M_p^{2/(3+p)} \, .  
\label{eqLamdaQG}
\end{equation}
We further assume $p>1$ and invoke SUSY
breaking via fluxes.\footnote{For details, see e.g., the discussion after
  Eq.~(3.40) in~\cite{Ferrara:1994kg}.}  This assumption leads to the gravitino conjecture
\begin{equation}
m_{3/2} = (\lambda_{3/2} \ m_{{\rm
    KK}_2})^{p} \ \frac{1}{M_s^{p-1}} \ ,
\label{gravconms}
\end{equation}
where we measure the gravitino mass not in units of the 4D Planck mass,
but in units of the string scale.

For a particular
example, we note that the modulino could be the fermionic partner of
the radion. Indeed, in the standard moduli stabilization by fluxes,
  all complex structure moduli and the dilaton are stabilized in a
  supersymmetric way while Kahler class moduli need an input from SUSY
  breaking. The radion is Kahler class and exists in a model
  independent fashion within the framework of the dark
  dimension. 

Next, using (\ref{m4}) with $k=1$ and $0.6 \alt \beta \alt 1.7$, it is
easily seen that the modulino mass
range shown in Fig.~\ref{fig:1} 
translates into $10^2 \alt m_{3/2}/{\rm TeV}  \alt 10^3$.
Substituting for $p=3$ and $\lambda_{3/2} =1/2$ into (\ref{gravconms}), we
obtain a gravitino mass in the ballpark: $m_{3/2} \sim 250~{\rm TeV}$. Finally, from 
(\ref{eqLamdaQG}) it follows that   
$M_s\sim M_* \sim 3  \times 10^{9}~{\rm GeV}$ and also that
$m_{{\rm KK}_2} \sim 2.6 \times 10^8~{\rm GeV}$.

Now, in line with our stated plan, we comment on two caveats of
the $k=1$ model discussed in this section:
\begin{itemize}[noitemsep,topsep=0pt]
\item We have no specific SUSY breaking mechanism in mind to supress
  the leading term of the modulino mass. Notwithstanding, by tuning
  the terms in the mass formula the leading term could be missing~\cite{Benakli:1997hb}.   
\item In (\ref{mkk1}) the cosmological constant requires its dominant
contribution from the Casimir energy, which scales as $m_{{\rm
    KK}_1}^4$. Additional contributions proportional to $M_*^4$, which
typically occur for high scale SUSY breaking must be absent or suppressed~\cite{Basile:2024lcz}. In
other words, the dark dimension subscribes to low scale SUSY breaking.
\end{itemize}

In summary, while the swampland program provides a suitable framework for
empirical calculations to determine the FPF sensitivity to modulino
oscillations, at the moment this scenario lacks of a concrete
theoretical model.

\section{Conclusions}

\label{sec:6}

We have shown that next generation LHC
experiments envisioned for the FPF would provide a unique opportunity to
probe stringy models with $M_s$ near the GUT scale by searching for signals
of neutrino-modulino oscillations. Models to be probed by FPF
experiments must have a SUSY breaking mechanism that isolates the {\it
  gravitino sector} from the {\it modulino sector}. We have shown that
SUSY breaking with sequestered gravity in gauge mediation provides an
explicit example in which the modulino mass is generated by radiative
corrections and is suppressed with respect to the gravitino mass.

We end with a pertinent observation. One major challenge in cosmology
is the well-known gravitino problem: gravitino decay leads to a huge
entropy production and if it takes place after big bang
nucleosynthesis (BBN) it could spoil its
predictions~\cite{Weinberg:1982zq,Ellis:1984er,Moroi:1993mb,Kohri:2005wn}. Now, sparticles
couple to visible matter only through gravitational
interactions, and so their couplings are Planck suppressed, yielding a gravitino decay
width $\Gamma_{3/2} \sim m_{3/2}^3/M_p^2$. It is straightforward to
verify that for all scenarios discussed in
this paper, the gravitino is cosmologically safe because its decay
width leads to a very short lifetime,
\begin{equation}
  \tau_{3/2} \sim 10^{-19}~{\rm s} \left(\frac{m_{3/2}}{10^{11}~{\rm GeV}}\right)^{-3} \ll t_{\rm BBN}    \,,
\end{equation}
where $t_{\rm BBN} \sim 600~{\rm s}$ is an estimate of the age of the universe at BBN.

\section*{Acknowledgements}

The work of L.A.A. is supported by the U.S. National Science
Foundation (NSF Grant PHY-2412679). I.A. is supported by the Second
Century Fund (C2F), Chulalongkorn University.  The work of D.L. is supported by the Origins
Excellence Cluster and by the German-Israel-Project (DIP) on Holography and the Swampland.

\end{document}